\documentclass[twocolumn,showpacs,preprintnumbers,amsmath,amssymb]{revtex4}
\usepackage{graphicx}
\usepackage{dcolumn}
\usepackage{bm}

\begin{document}
\title{Instability of two interacting, quasi-monochromatic  waves in 
shallow water}
\author{M. Onorato$^1$, D. Ambrosi$^2$, A. R. Osborne$^1$, M. Serio$^1$ }
\affiliation{$^1$Dip. di Fisica Generale, Universit\`{a} di Torino, via
Pietro Giuria 1, 10125 Torino, Italy\\
$^2$Dip. di Matematica, Politecnico di Torino, corso Duca degli Abruzzi 24,
10129 Torino, Italy}
\date{\today}
\begin{abstract} 
We study the nonlinear interactions of waves with 
a doubled-peaked power spectrum in shallow
water. The starting point is the prototypical equation for nonlinear 
uni-directional waves in shallow water, i.e. the Korteweg de Vries equation. 
Using a multiple-scale technique two defocusing coupled Nonlinear
Schr\"odinger equations are derived. We show  analytically that  
plane wave solutions of such a system can be unstable to 
small perturbations. This surprising result suggests 
the existence of a new energy exchange mechanism which could influence
the behaviour of ocean waves in shallow water.
\end{abstract}
\maketitle
%
%
The propagation of  multiple wave-train systems in shallow water
has historically received less attention than the propagation of
one single wave-train. Nevertheless, experimental studies carried out 
by Thompson \cite{THOM} in representative 
sites near the coasts of the United States reveal
that in $65 \%$ of the analyzed data, ocean wave spectra show two or more
separated peaks in the frequency domain  \cite{THOM}, \cite{SMITH92}.
In this framework, experimental work in the laboratory has been performed by 
Smith and Vincent \cite{SMITH92}. They propagated irregular wave trains with two distinct spectral peaks
 in a wave flume with 1:30 slope for different values
of the peak frequency and significant wave height. From a physical point of view, this condition mimics the interaction of two wave
regimes, a ``swell'' and a ``sea'', propagating in the same 
direction toward shore in shallow water. Their major
observation was a decay of the higher frequency peak along the flume. 
They hypothesized three possible explanations for such experimental results:
(i) resonant interactions among waves, (ii) bottom friction that acts 
differently for the two dominant wave peaks, (iii) breaking of the shorter waves
enhanced by the presence of the longer waves. More recently, using a higher order
Boussinesq model, Chen et al. \cite{CHEN97} have shown that 
inclusion of nonlinear 
interactions, without invoking bottom friction or wave breaking, is sufficient 
 to account for the decay of the high frequency peak 
.
Even though these numerical simulations of the Boussinesq equation
qualitatively reproduce the experimental results, the basic 
physical  mechanisms of interaction of wave trains
with double peaked spectra in shallow water are far from being completely understood.

In this paper  we discuss a fundamental instability that 
 occurs between two quasi-monochromatic interacting wave-trains.
Furthermore, the focus herein is not to attempt to model ocean waves but instead 
to study leading order effects using the simplest
 weakly nonlinear and dispersive model in shallow water, i.e. the Korteveg-de Vries (KdV)
 equation. A great deal of progress
 in understanding wave propagation in 
shallow water has been achieved by investigating the KdV equation, 
which can be considered as the basic 
weakly nonlinear model for unidirectional shallow water waves. 
The analytical properties of KdV (it is integrable by the 
Inverse Scattering Transform \cite{ABLSEG}) have improved 
basic  knowledge of the nonlinear dynamics of water waves \cite{MEI}, \cite{OSB1}, \cite{OSB2},\cite{OSB3}. In particular after the seminal work by Zabusky and Kruskal \cite{ZAB}, 
extensive studies have been carried out on the evolution 
of a {\it sine} wave in shallow water (see for example \cite{MEI}). 
However, less attention has been paid to the evolution 
of two monochromatic waves with different wave-numbers. 
The subject of this Communication is therefore an investigation of     
the basic nonlinear interaction that may take place between two
separated narrow-banded wave spectra in shallow water. To this aim, 
under suitable assumptions, using a multiple-scale technique, we 
derive from  the KdV equation a system of two 
Coupled Nonlinear Schr\"odinger (CNLS) equations.
Plane wave solutions of such a system are then studied analytically 
by a standard linear stability analysis, resulting in the presence of an instability 
region. 

The KdV equation can be formally derived from the Euler equations
for water waves \cite{MEI},\cite{WHIT} under the 
assumption that  waves are small (but finite amplitude) and long when  
compared with the water depth at rest. 
In a frame of reference moving with the velocity $c_0=\sqrt{g h}$, 
where $h$ is the water depth and $g$ is gravity acceleration, 
the KdV equation in nondimensional form reads:
\begin{equation} 
 \eta_t+ \mu\eta \eta_x + \lambda \eta_{xxx}=0.
\label{kdv}
\end{equation}
Here $\eta=\eta(x,t)$ is the free surface 
elevation, $x$ and $t$ are space and time variables; $\mu$ and $\lambda$
are the nonlinear and dispersive small parameters: 
$\mu=3 a /2 h$ and $\lambda=(h/l)^2 /6 $, with $a$ a characteristic 
wave amplitude and $l$ a characteristic wavelength.
We are interested in investigating the interaction of 
two waves, centered at nondimensional wave-numbers 
$k_1$ and $k_2$, propagating in the positive $x$ direction.
We consider the case of narrow-banded spectra, i.e. 
$\Delta k_i/k_i \ll 1$, with $i=1,2$, $\Delta k_i$ being 
the characteristic width of spectra around each peak. 
The approach used in our derivation of the CNLS equations
resembles the one used by
R. Grimshaw et al. \cite{GRIM} to derive a 
single higher order NLS equation starting from 
an extended KdV (cubic nonlinearity is included).
We introduce the following slow space and time variables $X=\epsilon x$ 
and $T=\epsilon t$, with $\epsilon \ll 1$, and perform a 
formal expansion of $\eta=\eta(x,t,X,T)$:
\begin{align}
\eta (x,&t,X,T) = \frac{\epsilon} {2}[A(X,T)e^{i \Theta_1}  + 
B(X,T)e^{i \Theta_2}+c.c.] \nonumber \\
& +   \frac{\epsilon^2} {2}[A_2(X,T)e^{2 i \Theta_1} +
B_2(X,T)e^{2 i \Theta_2}  \nonumber \\
& + C(X,T)e^{i (\Theta_1+\Theta_2)}+D(X,T)e^{i (\Theta_1-\Theta_2)}
+ c.c.]\nonumber \\
&+ \epsilon^2 M(X,T)+..., \label{expansion} 
\end{align}
where $\Theta_i=k_i x -\omega_i t$,
 $M(X,T)$ is a real quantity to be considered as a mean flow 
and $c.c.$ indicates complex conjugate.  From a physical 
point of view this representation corresponds to a double expansion
around wave numbers $k_1$ and $k_2$. 
 Note that terms at order $\epsilon^2$ include
 the complex envelopes $A_2$ and $B_2$ for the higher harmonics and 
 the complex envelopes $C$ and $D$ for the harmonics
given by the sum and difference of the fundamental wave-numbers.
  
After substituting the expansion (\ref{expansion}) into Equation 
(\ref{kdv}) and collecting terms for different harmonics,
 after some
lengthy but straightforward algebra,
we obtain a set of equations for the complex envelopes
and the mean flow $M$.
At order $\epsilon$, the equations for $A$ or $B$ provide 
the linear dispersion relation: $\omega_i=- \lambda k_i^3 $ ($i=1,2$).
For the second harmonics, at order $\epsilon^2$,  the following 
relations hold between complex 
envelopes:
\begin{align}
A_2&=\frac { \mu} { 12 \lambda k_1^2} A^2, \qquad
B_2=\frac { \mu} { 12 \lambda k_2^2} B^2, \label{higherharm} \\
C&= \frac { \mu} { 6 \lambda k_1 k_2} A B,\qquad
D= -\frac { \mu} { 6 \lambda k_1 k_2} A B^*.\nonumber
\end{align}
At order $\epsilon^3$,
 the mean flow $M$ is related to the 
envelopes $A$ and $B$ as follows:
\begin{equation}
M_T +  \frac {\mu} {4}   \bigg(|A|^2_X+|B|^2_X\bigg)=0.
\label{mean}
\end{equation}
Using Equation (\ref{higherharm}) and
neglecting terms of order higher than 
$\epsilon^3$, the equations for $A$ and $B$ then read:
\begin{align}
\epsilon^2(A_T-3 \lambda k_1^2 A_X)+ & i \epsilon^3 \bigg[3 \lambda k_1 A_{XX}
\label{carrier1}+ \\
& \bigg( \mu k_1 M + 
\frac{\mu^2}{24\lambda} \frac{1}{k_1} |A|^2 \bigg) A \bigg]=0,
\nonumber \\
\epsilon^2(B_T-3 \lambda k_2^2 B_X) +& i\epsilon^3 \bigg[ 3 \lambda k_2 B_{XX}
\label{carrier2} +\\
& \bigg( \mu k_2 M + 
\frac{ \mu^2} {24\lambda} \frac{1}{k_2} |B|^2 \bigg) B\bigg]=0.
\nonumber 
\end{align}
%
%
Note that the arguments used to derive 
Equations (\ref{carrier1}-\ref{carrier2})
are not valid when the two carrier 
waves $k_1$ and $k_2$ are equal 
(additional care in the expansion
should be taken to study the case of $k_1 \simeq k_2$).
Considering that at $\epsilon^2$ the following relations hold,
\begin{equation}
A_X \simeq \frac {1} {3 \lambda k_1^2} A_T, \qquad 
B_X \simeq \frac{1} {3 \lambda k_2^2} B_T,
\end{equation}
the mean flow can be directly related to the complex envelopes $A$ and $B$: integrating
Equation (\ref{mean}) in time we get  
\begin{equation}
M=-  \frac{\mu}{12 \lambda} \bigg( \frac {|A|^2} {k_1^2}  
+ \frac {|B|^2} {k_2^2} \bigg).
\label{mfcarrier}
\end{equation}
After substituting Equation (\ref{mfcarrier}) in 
(\ref{carrier1}), (\ref{carrier2}), and rescaling the variable as follows:
 $T=  \epsilon t/3 \lambda$, $X=\epsilon x$, 
$A'=A \epsilon k_1 \mu/(\lambda  6 \sqrt{2})$, 
$B'= B \epsilon k_2 \mu/(\lambda  6 \sqrt{2})$,
two defocusing Coupled Nonlinear Schr\"odinger equations
are obtained:
\begin{align}
A_t-  k_1^2 A_x+  i  k_1 A_{xx} - i\frac {1 }{k_1^3}{|A|^2A} - 
2 i  \frac {k_1} {k_2^4} |B|^2A =0,
\label{COUP11} \\
B_t-  k_2^2 B_x+  i  k_2 B_{xx}
- i\frac {1 }{k_2^3} |B|^2B - 2 i  \frac {k_2} {k_1^4}|A|^2B  =0,
\label{COUP22}
\end{align}
where we have dropped the primes for simplicity. 
%
In these new variables the complex envelopes 
appearing in (\ref{COUP11}-\ref{COUP22})
are proportional to the Ursell number \cite{MEI},\cite{OSB1} 
($U_r=ka/(kh)^3$),
therefore to the nonlinearity 
of the system. 
If we set $B=0$, the system reduces identically to the 
well known defocusing Nonlinear 
Schroedinger equation for the complex envelope $A$,
originally  derived from the Euler equations 
in the limit of shallow water, for narrow-banded 
spectra and small steepness \cite{HAONO}.
The single defocusing NLS equation has also been obtained
 from the KdV equation 
in \cite{ZAKHK}, see also \cite{TRACY}.
For a single wave train the expansion (\ref{expansion})
is nothing but the second-order 
modulated Stokes wave, i.e. the Stokes series in 
shallow water. It is well known 
that the plane wave solution of the NLS equation in shallow 
water (defocusing NLS) is stable to side-band perturbations. This 
contrasts to the Benjamin-Feir instability of the  focusing NLS that describes the propagation of waves in deep water. 

Systems of equations similar to (\ref{COUP11}-\ref{COUP22}) 
are not new  in various fields of physics. For example, focusing CNLS equations 
have been derived in \cite{Menyuk87}, \cite{Menyuk89} for optical fibers.
For particular values of the coefficients 
of the nonlinear terms in (\ref{COUP11}-\ref{COUP22}), 
the integrable Manakov system is recovered \cite{MAN}
(see also \cite{FOREST}).
Concerning water waves, 
focusing CNLS equations have been
 derived for the first time by Roskes \cite{ROSK} 
in infinite water depth.  CNLS systems describing
the interaction of two counterpropagating waves 
are discussed in \cite{OKA}, \cite{KOP}.
Unlike the work by Roskes, the attention here is focused 
on the  shallow water limit, i.e. 
in the case where the system reduces to
two {\it defocusing} NLS equations.

We now consider the linear stability analysis of
the system (\ref{COUP11}-\ref{COUP22}).
Consider a plane wave solution of CNLS of the form:
\begin{eqnarray}
A=\tilde {A} e^{ -i \omega_A t}, \qquad      
B=\tilde {B} e^{-i  \omega_B t},
\label{planesolution}
\end{eqnarray}
with 
\begin{align}
\omega_A=-(|\tilde {A}|^2/k_1^3+2|\tilde  {B}|^2 k_1/k_2^4), \\   
\omega_B=-(|\tilde {B}|^2/k_2^3+2 |\tilde {A}|^2 k_2/k_1^4),
\end{align}
i.e. the nonlinear frequency correction in the Stokes
expansion. The solutions (\ref{planesolution}) are then perturbed as follows:
\begin{eqnarray}
A=\tilde {A}(1+a) e^{-i (\omega_A t + \alpha )}, \nonumber\\
B=\tilde {A} (1+b) e^{-i ( \omega_B t + \beta)},
\label{perturbation}
\end{eqnarray}
where $a$, $b$, $\alpha$ and $\beta$ 
are small perturbations in amplitude and phase.
We substitute relations  
(\ref{perturbation}) into the CNLS equations and obtain a 
system for $a$, $b$, $\alpha$ and $\beta$; 
after linearization, the standard Fourier technique is used for
solving it:
\begin{eqnarray}
a=\tilde {a} e^{i (K x -\Omega t)}, \qquad
b=\tilde {b} e^{i (K x -\Omega t)}, \nonumber\\
\alpha=\tilde {\alpha}e^{i (K x -\Omega t)},\qquad
\beta=\tilde {\beta} e^{i (K x -\Omega t)},
\label{perturbation2}
\end{eqnarray}
where $\tilde {a}$,$\tilde {b}$, $\tilde {\alpha}$ and $\tilde {\beta}$ are
constant, $\Omega=\Omega(K)$ is the dispersion relation for 
perturbed wave-number $K$.
After some algebra, the dispersion relation results in a
 fourth-order polynomial function in the variable $\Omega$:
\begin{align}
k_1^2 k_2^2 \bigg((k_1^2 K +\Omega)^2- k_1^2   K^4 
 -2  |\tilde{A}|^2 K^2/k_1^2\bigg)\times& \label{disppert} \\
\bigg((k_2^2 K +\Omega)^2- k_2^2   K^4   
- 2 |\tilde{B}|^2 K^2/k_2^2\bigg)=& \nonumber\\
16  |\tilde{A}|^2|\tilde{B}|^2 K^4&. \nonumber
\end{align}
Equation (\ref{disppert}) provides the dispersion 
relation for the perturbation: 
complex roots, which are functions of the parameters
$k_1$, $k_2$, $\tilde{A}$, $\tilde{B}$ and $K$, originate instability. 
Note that if we set one of the two
amplitudes equal to zero (for example $\tilde{B}=0$), the right hand-side term 
in (\ref{disppert}) vanishes 
and roots can be found easily. In this case the roots are always real:
a single monochromatic wave in shallow water is
stable to side-band perturbations.
Equation (\ref{disppert}) has been investigated 
by applying the exact
formula for the solution of a fourth-order polynomial equation. Complex roots
are then found. Results are presented in the following way: we fix the 
nondimensional wave-number $k_1=1$ (this choice is natural if
we take the characteristic length $l$ to be $1/k$, with $k$ the dimensional
wave-number of the first peak in the spectrum);  amplitudes $\tilde{A}$ and 
$\tilde{B}$
are then selected in order to fix the nonlinearity of the 
initial conditions.
\begin{figure}
\includegraphics{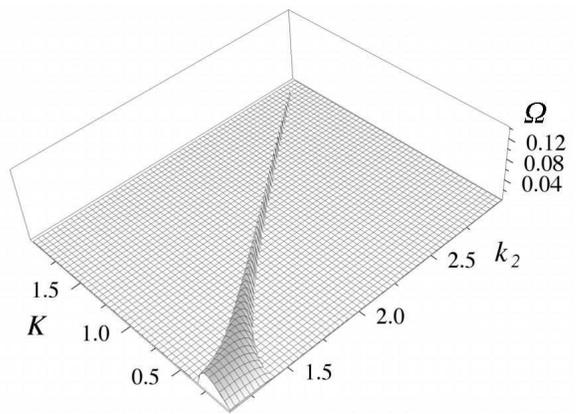} 
\caption{\label{figura1} Growth-rate as a function of nondimensional 
wave-number $k_2$ and perturbation $K$ for $\tilde{A}=\tilde{B}=0.2$
 and $k_1=1$. Axes
are in nondimensional units.}
\end{figure}
\begin{figure}
\includegraphics{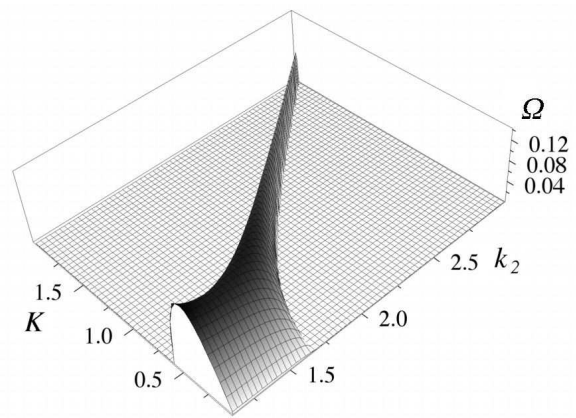} 
\caption{\label{figura2} Growth-rate as a function of nondimensional 
wave-number $k_2$ and perturbation $K$ for 
$\tilde{A}=\tilde{B}=0.4$ and $k_1=1$. Axes
are in nondimensional units.}
\end{figure}
 We then compute the largest 
imaginary solution of the polynomial function $\Omega$ and plot 
it as a surface for different values 
of $k_2$ and perturbation $K$; as $k_2$ is 
increased, the amplitude $\tilde{B}$ is kept constant. In dimensional
variables this would correspond to increasing the wave-number $k_2$
and at the same time maintaining constant its Ursell number. 

In Fig. \ref{figura1} we show the instability region
for $\tilde{A}=0.2$, $\tilde{B}=0.2$. The plot clearly 
exhibits an unstable region
with a growth-rate different from zero.
Note that on the horizontal axes $k_2$ starts from values slightly larger than
1, because, as has been previously 
stated, the CNLS system that we have derived 
is not valid when $k_2 \simeq k_1$.
In Fig. \ref{figura2} we show the same diagram 
as in Fig. \ref{figura1}, with larger values of the amplitudes
$\tilde{A}=0.4$ and $\tilde{B}=0.4$. The instability and the growth-rate have
 now both increased. 
 We recall that 
the amplitudes $\tilde{A}$ and $\tilde{B}$ have been scaled by a combination
of the steepness, of the nonlinear and dispersive parameters,
 respectivelly $\mu$ and $\lambda$, of the KdV equation. 
 From a physical point of view,
increasing the amplitude $\tilde{A}$ or $\tilde{B}$ implies an increase
of the nonlinearity of the wave system that can be achieved for example 
also by a decrease in the water depth. As the the water depth
decreases the instability region and the growth-rate 
increase and therefore, double-peak spectra are more likely to
evolve naturally into single-peak spectra which are stable in shallow
water. Results can be summarized as follows: in shallow water
the dynamics of two wave trains can be unstable; 
as expected the growth-rate and the size of the instability region 
depend on the nonlinearity of the system.

The derived CNLS system represents a very crude simplification of the real 
problem: effects related to a non constant water depth could be considered;
directionality could also be simply included by applying the multiple-scale method  to the Kadomtsev-Petviashvili 
equation \cite{ABLSEG} (an extension of the KdV equation
that includes the dynamics
of transverse perturbations); higher order effects
could also be investigated.
 
We believe that the present work offers  new perspectives
for understanding the dynamics of double-peaked spectra in shallow water.
Accurate comparison with experiments 
and numerical simulations are 
definitely needed in order to validate the obtained 
theoretical results.
As a final remark we would like to stress that we have 
started the derivation of the CNLS equations from
the KdV; the Inverse Scattering theory furnishes
a unique method for investigating all its solutions. 
It would be therefore interesting to interpret the unstable solutions
of the CNLS equations in terms of the Inverse Scattering modes
for the KdV equation.

{\bf Acknowledgements}
M.O. is grateful to J.M. Smith for pointing
out the problem. 
This work was supported by  the O.N.R., USA.
Torino University funds (60 \%) are also acknowledged.
D. Resio is acknowledged for support 
and valuable discussions. 
%
%
%
%
%

\end{document}